# Influence of substrate type on transport properties of superconducting $FeSe_{0.5}Te_{0.5}$ thin films

Feifei Yuan[1,2], Kazumasa Iida[2,3†], Marco Langer[2,4], Jens Hänisch[2,4], Ataru Ichinose[5], Ichiro Tsukada[5], Alberto Sala[6], Marina Putti[6], Ruben Hühne[2], Ludwig Schultz[2] and Zhixiang Shi[1‡]

[1]Department of Physics and Key Laboratory of MEMS of the Ministry of Education, Southeast University, Nanjing 211189, People's Republic of China

[2]Institute for Metallic Materials, IFW Dresden, D-01171 Dresden, Germany

[3]Department of Crystalline Materials Science Graduate School of Engineering, Nagoya University, Nagoya 464-8603, Japan

[4] Karlsruhe Institute of Technology, Institute for Technical Physics, D-76344 Eggenstein-Leopoldshafen, Germany

[5]Central Research Institute of Electric Power Industry, 2-6-1 Nagasaka, Yokosuka, Kanagawa 240-0196, Japan

[6]Dipartimento di Fisica, Università di Genova and CNR-SPIN, Via Dodecaneso 33, 16146 Genova, Italy

Email: iida@nuap.nagoya-u.ac.jp

zxshi@seu.edu.cn

## Abstract

$FeSe_{0.5}Te_{0.5}$ thin films were grown by pulsed laser deposition on $CaF_2$, $LaAlO_3$ and MgO substrates and structurally and electro-magnetically characterized in order to study the influence of the substrate on their transport properties. The in-plane lattice mismatch between $FeSe_{0.5}Te_{0.5}$ bulk and the substrates shows no influence on the lattice parameters of the films, whereas the type of substrates affects the crystalline quality of the films and, therefore, the superconducting properties. The film on MgO showed an extra peak in the angular dependence of critical current density $J_c(\theta)$ at $\theta = 180°$ ($H \parallel c$), which arises from $c$-axis defects as confirmed by transmission electron microscopy. In contrast, no $J_c(\theta)$ peaks for $H \parallel c$ were observed in films on $CaF_2$ and $LaAlO_3$. $J_c(\theta)$ can be scaled successfully for both films without $c$-axis correlated defects by the anisotropic Ginzburg-Landau (AGL) approach with appropriate anisotropy ratio $\gamma_J$. The scaling parameter $\gamma_J$ is decreasing with decreasing temperature, which is different from what we observed in $FeSe_{0.5}Te_{0.5}$ films on Fe-buffered MgO substrates.

**Keywords**

## 1. Introduction

After the discovery of superconductivity at 26 K in the iron oxypnictide, LaFeAs(O,F) [1], extensive research on iron-based superconductors has been carried out. The novel iron-based superconductors are categorized mainly in four systems with different crystal structures: "1111" [$RE$FeAs(O,F), ($RE$: rare earth elements)] [1], "122" [$AE$Fe$_2$As$_2$, ($AE$: alkaline earth elements)] [2], "111" (LiFeAs) [3] and "11" (Fe-chalcogenides) [4]. FeSe as a member of the 11 family has a superconducting transition temperature ($T_c$) of around 8 K, which can be enhanced up to 37 K with the application of external pressure [5, 6]. The partial substitution of Se by Te also leads to an enhancement of $T_c$ to 14 K [7]. Due to its simple crystal structure and the less toxic nature, the iron chalcogenides are considered to be suitable for exploring the mechanism of superconductivity as well as for applications. Compared to single crystals, thin films are suitable for investigating transport properties and superconducting electronics applications thanks to their geometry. Fe(Se,Te) films have already been grown on various substrates [8-14], applying different lattice or thermal expansion mismatch between films and substrates. However, the role of the substrate for the superconducting properties is still not clear. The maximum $T_c$ of 21 K has been obtained by Bellingeri *et al.* for films deposited on LaAlO$_3$ (LAO) substrates [8], which they attributed to the compressive strain due to the growth mode. On the other hand, Imai *et al.* showed that the superconducting properties are not well correlated to the lattice mismatch of the substrates but rather connected to the cell constants of the films [9, 10]. Furthermore, films on SrTiO$_3$ show a peak in the angular dependence of $J_c$ for $H \parallel c$ [11-13]. Similar $c$-axis peaks in $J_c(\theta)$ have been observed for films deposited on $CaF_2$ substrate [14]. However, no additional peaks have been observed in $FeSe_{0.5}Te_{0.5}$ films on LAO [11] and Fe-buffered MgO substrates [15-16]. In this work, $FeSe_{0.5}Te_{0.5}$ thin films were deposited on $CaF_2$, LAO and MgO single crystalline substrates in order to further elucidate the role of the substrate for the superconducting properties.

## 2. Experiment details

$FeSe_{0.5}Te_{0.5}$ films were prepared on $CaF_2$ ($a/\sqrt{2} = 0.386$ nm), LAO ($a = 0.379$ nm) and MgO ($a = 0.4212$ nm) (001)-oriented cubic single crystalline substrates by pulsed laser deposition (PLD) with a KrF excimer laser (wavelength: 248 nm, repetition rate: 7 Hz) under ultrahigh vacuum (UHV) conditions with a background pressure of $10^{-9}$ mbar [15-16]. The substrate temperature was fixed to 400℃ during deposition.

Structural properties of the films were investigated by x-ray diffraction (XRD) in $\theta$-$2\theta$ geometry at a Bruker D8 Advance with Co-$K_\alpha$ radiation and at a texture goniometer Phillips X'pert with Cu-$K_\alpha$ radiation. The $c$ lattice parameters were calculated from $\theta$-$2\theta$ scans using the Nelson Riley function. The lattice parameters $a$ were derived from reciprocal space maps measured in a Panalytical X'pert Pro system. Transmission electron microscope (TEM) investigations of the films have been performed on a JEOL JEM-2100F microscope. The film thicknesses were determined from cross-sectional views, for which the films were cut by a focused ion beam technique (FEI Helios 600i) and estimated to be 140 nm, 180 nm and 250 nm for the film on MgO, $CaF_2$ and LAO, respectively.

For transport measurements, microbridges of 100 μm width and 4.1 mm length were fabricated by ion beam etching after a photolithographic process. Silver paint was employed for electrical contacts. Electrical transport properties were measured in a Physical Property Measurement System [(PPMS) Quantum Design] by a standard four-probe method. The upper critical field $H_{c2}$ was defined with a criterion of 90% of the normal state resistance $R_n = R(22$ K), the irreversibility field $H_{irr}$ was defined with a criterion of 1% $R_n$. $J_c$ was defined with a criterion of $1\mu$V cm$^{-1}$. In the angular dependent $J_c$ measurements, the magnetic field $H$ was applied in maximum Lorentz force configuration ($H \perp J$, where $J$ is the current density) at an angle $\theta$ from the $c$ axis.

### 3. Results and discussion

Figure 1 summarizes the structural characterization of the prepared samples by XRD. In figure 1 (a), only sharp 00$l$ peaks of the FeSe$_{0.5}$Te$_{0.5}$ films and the respective substrates are present with exception of a minute unidentified peak for the film on LAO, indicating a high phase purity with $c$-axis alignment of the films. The (001) rocking curves of the three films, figure 1 (b), show a similar out-of-plane orientation spread of around 0.8 ° full width at half maximum, FWHM. The rocking curves show a slight asymmetry, especially for LAO and MgO, which indicates mosaicity and the incorporation of grain boundary (GB) networks. The $\varphi$-scans of the 101 reflection, figure 1 (c), show fourfold symmetry with peaks at every 90 °, indicating that the films are epitaxially grown with a cube-on-cube relationship towards the substrates in case of LAO and MgO. The film on CaF$_2$ also has four-fold symmetry; however, the basal plane of the FeSe$_{0.5}$Te$_{0.5}$ film is rotated by 45 ° with respect to the substrate. The respective average FWHM of the $\varphi$-scan peaks (i.e., 4 peaks) are 0.56 ° and 0.75 ° for the film on CaF$_2$ and LAO. The largest value of $\Delta\varphi = 2.35$ ° is observed for the film deposited on MgO.

These results indicate that the crystalline quality of the film deposited on MgO is inferior to the other films. The larger lattice mismatch between the film and MgO substrate might be the reason for the poorer crystallinity.

The lattice parameters of the films are summarized in table I. The lattice constants $c$ and $a$ of all films are smaller than those of FeSe$_{0.5}$Te$_{0.5}$ single crystals ($a$ = 0.3801 nm, $c$ = 0.6034 nm) [7], but in agreement with the reported values for thin films [17-20]. The film grown on CaF$_2$ has the largest $c$ value and the smallest $a$ value, which might originate from a difference in thermal expansion coefficient between the substrate and superconducting layer. In general, there is no clear correlation between the lattice parameters of the films and those of the substrates. In particular, the $a$-axis length for the LAO and MgO substrates are 0.379 nm and 0.4212 nm, respectively, whereas the $a$-axis parameter of the grown films is almost identical, indicating that the lattice constants $a$ of the films are not correlated to the in-plane lattice parameters of the substrates (Table I).

The cross-sectional TEM images for the FeSe$_{0.5}$Te$_{0.5}$ thin films are shown in figure 2. A bright area with 5 nm in width is observed at the interface between the CaF$_2$ substrate and the superconducting film [figure 2 (a)], similar to observations in Ref. [21]. This is presumably a reaction layer between film and substrate. In comparison, the film-substrate interface is smooth and clean for the layers on LAO and MgO [figure 2 (b) and 2 (c)]. For the film on CaF$_2$, no granular structure or large extended defects are observed throughout the whole film. However, the lattice seems to be disturbed in small regions [figure 2 (d)]. Inside the film on LAO, on the other hand, dark islands (probably small grains) disperse and modulate the structure, figure 2 (e). In stark contracts to these two films, defects parallel to the $c$ axis are found in the film on MgO, shown in figure 2 (c). Additionally, areas of different intensity appear in the film, indicating a slight crystal distortion or crystal rotation shown in figure 2 (f).

The value of $T_c$ is not related to the films' lattice constant, but rather to the type of substrates and the crystallinity of films. Even though the films on LAO and MgO show the same $a$-axis parameter, their resistive $T_c$ values differ considerably, figure 3. The lowest $T_c$ of 16.3 K is measured for the film on MgO with extended defects and the poorest degree of texture. The $T_c$ of the film on LAO is 18.0 K, which is 1.7 K higher. The film on CaF$_2$, showing the best crystalline quality with $\Delta\omega \sim 0.8°$ and $\Delta\varphi \sim 0.56°$, exhibits the highest $T_c$ (19.1 K), which is about 5 K higher than the bulk value [7]. The epitaxial compressive strain induced by the CaF$_2$ substrate due to thermal misfit might be crucial for such a high $T_c$ value [22].

The superconducting transition shifts to lower temperatures in applied magnetic fields, as exhibited in the inset of figure 3 for the film on CaF$_2$ for $H \parallel ab$. The decrease of the transition is less than 1 K even for a field of 9 T, indicating high upper critical fields $H_{c2}$ at low temperatures. Figure 4 shows $H_{c2}$ and $H_{irr}$ for all films as a function of the reduced temperature $t = T/T_c$ for fields parallel and perpendicular to the $c$ axis. All films have a high slope near $T_c$, with $\left|\frac{d\mu_0 H_{c2}}{dT}\right|_{T_c}$ ranging from 8.6 to 13.3 T/K for $H \parallel c$. The three films have a similar $H_{c2}$ slope for magnetic fields perpendicular to $c$ axis of around 36 T/K. According to the relation $\xi_{ab}^2 = \Phi_0/(2\pi * 0.691 * T_c * S^c)$ [24], where $\Phi_0 = 2.07 \times 10^{-15}$ Wb and $S^c = dH_{c2}^c/dT$ is the slope of $H_{c2}^c$ near $T_c$ and $\xi_c/\xi_{ab} = S^{ab}/S^c$, this corresponds to coherence lengths $\xi_c$ ranging from 0.42 nm to 0.53 nm. This $\xi_c$ values are larger than half the interlayer spacing ($c/2$), and 2D effects, such as intrinsic pinning, are not expected.

The magnetic field dependencies of $J_c$ are compared at a reduced temperature $t = 0.5$, shown in figure 5. The $J_c$ values for $H \parallel ab$ are always higher than those for $H \parallel c$. Again, the film on $CaF_2$ exhibits the highest $J_c$ values in both crystallographic directions. A $J_c$ as high as $10^5$ A/cm$^2$ at 9 T ($H \parallel ab$) was measured, which is comparable to the values reported in Refs. [21, 23]. The film on MgO has the lowest $J_c$ value, presumably due to the poor crystalline quality.

The angular dependence of $J_c$ of all films in various magnetic field strengths, shown for 4 K in figure 6, exhibit a broad maximum positioned at $\theta = 90°$ owing to the mass anisotropy of the material. For the films on LAO and $CaF_2$, no additional peaks at $\theta = 180°$ were observed in the whole range of magnetic fields and temperatures. In the case of the film grown on MgO, an additional peak at $\theta = 180°$ (i.e. $c$-axis peak) is visible for all magnetic fields, as displayed in the inset of figure 6(c). This peak is related to correlated defects confirmed by TEM acting as effective pinning centers parallel to the $c$-axis [11-14].

$J_c$ can be scaled with an effective magnetic field $H_{eff}$, [ $H_{eff} = H \cdot \varepsilon$, $\varepsilon = \sqrt{\cos^2\theta + \gamma_J^{-2}\sin^2\theta}$ ], where $\gamma_J$ is the scaling parameter and related to the mass anisotropy ratio [25-29]. The scaling behavior of $J_c$ as a function of $H_{eff}$ at various temperatures is shown for the films on $CaF_2$ and LAO in figure 7. All data except for those in the vicinity of $H \parallel ab$ collapse onto the experimental curves $J_c$ ($H \parallel c$) with $\gamma_J$ values between 2 and 3. $\gamma_J$ shows the same temperature dependence as the upper critical field anisotropy $\gamma_{Hc2}$ of films and single crystals [30]. According to the two band theory [31], the existence of impurities, causing strong intraband scattering and negligible interband scattering, results in the decrease of $\gamma_{Hc2}$ with decreasing temperature. The strong increase of $\gamma_{Hc2}$ for $T$ close to $T_c$ is presumably caused by the negative curvature of $H_{c2}^{ab}$ due to the paramagnetic limitation, as determined for single crystals [32]. $\gamma_J$ shifts to larger values compared with $\gamma_{Hc2}$ due to the presence of strong pinning centers, a similar observation as in cuprates [33], where $\gamma_J$ is lowered by extended isotropic defects. In contrast to the films of this study, $\gamma_J$ of films on Fe-buffered MgO with a much shorter coherence length $\xi_c$ [16], follows the trend and magnitude of the penetration depth anisotropy $\gamma_\lambda$ instead. It is difficult and not reasonable to scale $J_c$ of the film on bare MgO, due to the extra $c$-axis peak in $J_c$ related to the line defects as confirmed by TEM images.

## 4. Conclusions

FeSe$_{0.5}$Te$_{0.5}$ thin films were deposited on a variety of substrates by PLD. A systematic study of the substrate dependence on the superconducting properties of the films has been carried out. The lattice parameters of the films are not related to the lattice misfit between the substrates and films. Although both lattice parameters $a$ and $c$ are comparable for the films on MgO and LAO, the $T_c$ is about 2 K lower for the film on MgO compared to the one on LAO. The crystalline quality and epitaxy of the films may play a critical role for the superconducting properties. A $c$-axis $J_c$ peak was observed for the film deposited on MgO substrate due to the line defects parallel to $c$ axis found by TEM investigations. $J_c(\theta)$ of the films on CaF$_2$ and LAO were scaled using the anisotropic Ginzburg-Landau approach with appropriate scaling parameters $\gamma_J$. $\gamma_J$ is decreasing with decreasing temperature, which is attributed to multi-band superconductivity.

**Acknowledgement**

The authors thank V. Grinenko for fruitful discussion, I Mönch for help with microbridges preparation, J. Scheiter for help with FIB cuts and TEM lamellae preparation, and S. Molatta, M. Kühnel and U. Besold for technical support. The research leading to these results has received funding from European Union's Seventh Framework Programme (FP7/2007-2013) under grant agreement numbers 283141 (IRON-SEA) and 283204 (SUPER-IRON) the National Basic Research Program of China (973 Program, Grant No. 2011CBA00105). The work was partly supported by the Strategic International Collaborative Research Program (SICORP), Japan Science and Technology Agency, and National Science Foundation of China (Grant No. NSFC-U1432135).

**Reference**

[1] Kamihara Y, Watanabe T, Hirano M and Hosono H 2008 *J. Am. Chem. Soc.* **130** 3296.
[2] Rotter M, Tegel M, Johrendt D 2008 *Phys. Rev. Lett.* **101** 107006.
[3] Wang X C, Liu Q Q, Lv Y X, Gao W B, Yang L X, Yu R C, Li F Y and Jin C Q 2008 *Solid State Commun.* **148** 538-540.
[4] Hsu F C, Luo J Y, Yeh K W, Chen T K, Huang T W, Mu P M, Lee Y C, Huang Y L, Chu Y Y, Yan D C and Wu M K 2008 *Proc. Natl. Acad. Sci. USA* **105** 14262.
[5] Margadonna S, Takabayashi Y, Ohishi Y, Mizuguchi Y, Takano Y, Kagayama T, Nakagama T, Takata M and Prassides K 2009 *Phys. Rev. B* **80** 064506.
[6] Medvedev S, McQueen T M, Troyan I A, Palasyuk T, Eremets M I, Cava R J, Naghavi S, Casper F, Ksenofontov V, Wortmann G and Felser C 2009 *Nat. Mater.* **8** 630
[7] Sales B C, Sefat A S, McGuire M A, Jin R Y, Mandrus D and Mozharivskyj Y 2009 *Phys. Rev. B* **79** 094521.
[8] Bellingeri E, Pallecchi I, Buzio R, Gerbi A, Marrè D, Cimberle M R, Tropeano M, Putti M, Palenzona A and Ferdeghini C 2010 *Appl. Phys. Lett.* **96** 102512.
[9] Imai Y, Akiike T, Hanawa M, Tsukada I, Ichinose A, Maeda A, Hikage T, Kawaguchi T and Ikuta H 2010 *Appl. Phys. Express* **3** 043102.
[10] Hanawa M, Ichinose A, Komiya S, Tsukada I, Akiike T, Imai Y, Hikage T, Kawaguchi T, Ikuta H and Maeda A 2011 *Jpn. J. Appl. Phys.* **50** 053101.
[11] Bellingeri E, Kawale S, Braccini V, Buzio R, Gerbi A, Martinelli A, Putti M, Pallecchi I, Balestrino G, Tebano A and Ferdeghini C 2012 *Supercond. Sci. Technol.* **25** 084022.
[12] Bellingeri E, Kawale S, Pallecchi I, Gerbi A, Buzio R, Braccini V, Palenzona A, Putti M, Adamo M, Sarnelli E and Ferdeghini C 2012 *Appl. Phys. Lett.* **100** 082601.
[13] Braccini V, Kawale S, Reich E, Bellingeri E, Pellegrino L, Sala A, Putti M, Higashikawa K, Kiss T, Holzapfel B, Ferdeghini C 2013 *Appl. Phys. Lett.* **103** 172601.

[14] Mele P, Matsumoto K, Fujita K, Yoshida Y, Kiss T, Ichinose A and Mukaida M 2012 *Supercond. Sci. Technol.* **25** 084021.

[15] Iida K, Hänisch J, Schulze M, Aswartham S, Wurmehl S, Büchner B, Schultz L and Holzapfel B 2011 *Appl. Phys. Lett.* **99** 202503.

[16] Iida K, Hänisch J, Reich E, Kurth F, Hühne R, Schultz L, Holzapfel B, Ichinose A, Hanawa M, Tsukada I, Schulze M, Aswartham S, Wurmehl S and Büchner B 2013 *Phys. Rev. B* **87** 1045100.

[17] Kumary T G, Baisnab D K, Janaki J, Mani A, Satya A T, Sarguna R M, Ajikumar P K, Tyagi A K and Bharathi A 2009 *Supercond. Sci. Technol.* **22** 095018.

[18] Si W D, Lin Z W, Jie Q, Yin W G, Zhou J, Gu G D, Johnson P D and Li Q 2009 *Appl. Phys. Lett.* **95** 052504.

[19] Huang S X, Chien C L, Thampy V and Broholm C 2010 *Phys. Rev. Lett.* **104** 217002.

[20] Zhuang J C, Yeoh W K, Cui X Y, Kim J H, Shi D Q, Shi Z X, Ringer S P, Wang X L and Dou S X 2014 *Appl. Phys. Lett.* **104** 262601.

[21] Tsukada I, Hanawa M, Akiike T, Nabeshima F, Imai Y, Ichinose A, Komiya S, Hikage T, Kawaguchi T, Ikuta H and Maeda A 2011 *Appl. Phys. Express* **4** 053101.

[22] Ichinose A, Tsukada I, Nabeshima F, Imai Y, Maeda A, Kurth F, Holzapfel B, Iida K, Ueda S and Natio M 2014 *Appl. Phys. Lett.* **104** 122603.

[23] Kawale S, Bellingeri E, Braccini V, Pallecchi I, Putti M, Grimaldi G, Leo A, Guarino A, Nigro A and Ferdeghini C 2013 , *IEEE Trans. Appl. Supercond.* **23** 7500704.

[24] Werthamer N R, Helfand E and Hohenberg P C 1966 *Phys. Rev.* **147** 295.

[25] Blatter G, Geshkenbein V B and Larkin A I 1992 *Phys. Rev. Lett.* **68** 875.

[26] Iida K, Hänisch J, Thersleff T, Kurth F, Kidszun M, Haindl S, Hühne R, Schultz L and Holzapfel B 2010 *Phys. Rev. B* **81** 100507.

[27] Iida K, Haindl S, Thersleff T, Hänisch J, Kurth F, Kidszun M, Hühne R, Mönch I, Schultz L, Holzapfel B and Heller R 2010 *Appl. Phys. Lett.* **97** 172507.

[28] Hänisch J, Iida K, Haindl S, Kurth F, Kauffmann A, Kidszun M, Thersleff T, Freudenberger J, Schultz L and Holzapfel B 2011 *IEEE Trans. Appl. Supercond.* **21** 2887.

[29] Kidszun M, Haindl S, Thersleff T, Hänisch J, Kauffmann A, Iida K, Freudenberger J, Schultz L and Holzapfel B 2011 *Phys. Rev. Lett.* **106** 137001.

[30] Audouard A, Drigo L, Duc F, Fabrèges X, Bosseaux L and Toulemonde P 2014 *J. Phys.: Condens. Matter.* **26** 185701.

[31] Gurevich A 2003 *Phys. Rev. B* **67** 184515.

[32] Klein T, Braithwaite D, Demuer A, Knafo W, Lapertot G, Marcenat C, Rodière P, Sheikin I, Strobel P, Sulpice A and Toulemonde P 2010 *Phys. Rev. B* **82** 184506.

[33] Gutiérrez J, Llordés A, Gázquez J, Gibert M, Romà N, Ricart S, Pomar A, Sandiumenge F, Mestres N, Puig T and Obradors X 2007 *Nat. Mater.* **6** 367-373.

Table I Structural and superconducting properties of the films on different substrates

| Substrate | $a_{sub}$ (nm) | $\triangle \omega$ (°) | $\triangle \varphi$ (°) | $c$ (nm) | $a$ (nm) | $c/a$ | Thickness (nm) | $T_c$ (K) | $\left\|\dfrac{d\mu_0 H_{c2}^c}{dT}\right\|_{T_c}$ (T K$^{-1}$) | $\xi_c$ (nm) |
|---|---|---|---|---|---|---|---|---|---|---|
| CaF$_2$ | 0.386 | 0.80 | 0.56 | 0.597 | 0.374 | 1.596 | 180 | 19.1 | 13.2 | 0.50 |
| LAO | 0.379 | 0.70 | 0.75 | 0.592 | 0.377 | 1.570 | 250 | 18.0 | 8.6 | 0.42 |
| MgO | 0.421 | 0.85 | 2.35 | 0.595 | 0.377 | 1.578 | 140 | 16.3 | 13.3 | 0.53 |

**Figure Captions**

Figure 1. Structural properties of $FeSe_{0.5}Te_{0.5}$ films deposited on $CaF_2$, LAO and MgO substrates. (a) $\theta/2\theta$ scan. The asterisk represent an unidentified peak. (b) Rocking curve for the (001) reflection of the films. (c) $\varphi$ scans using (101) reflection.

Figure 2. Cross-sectional TEM images of (a)-(c) the interface between $FeSe_{0.5}Te_{0.5}$ thin film and substrate and (d)-(f) films on $CaF_2$, LAO and MgO substrates.

Figure 3. Resistive superconducting transition of the films on different substrates in zero magnetic field. The inset shows the resistive transitions of the film on $CaF_2$ measured in magnetic fields up to 9 T for $H \parallel ab$.

Figure 4. The upper critical fields $H_{c2}$ and irreversibility field $H_{irr}$ as a function of the reduced temperature $T/T_c$ for field parallel and perpendicular to the $c$ axis.

Figure 5. Magnetic field dependence of critical current density at $T/T_c = 0.5$ for (a) $H \parallel c$ and (b) $H \parallel ab$.

Figure 6. Angular dependence of the critical current density measured at 4 K in various magnetic field strengths for the films on (a) $CaF_2$, (b) LAO and (c) MgO substrates. The inset for the film on MgO shows the enlarged view in the angular range from 120 ° to 220 °.

Figure 7. The scaling behavior of $J_c(\theta)$ as a function of $H_{eff}$ at various temperatures for thin

films on (a) $CaF_2$ and (b) LAO. The solid lines represent the measured $J_c(H \parallel c)$ values. (c) Temperature dependence of the anisotropy ratio $\gamma_J$ obtained by the AGL scaling. $\gamma_J$ values for a $FeSe_{0.5}Te_{0.5}$ thin film on Fe-buffered MgO substrate (Ref. [16]) and the $H_{c2}$ anisotropy parameter $\gamma_{Hc2}$ for $FeSe_{0.5}Te_{0.5}$ single crystal (Ref. [30]) are also plotted for comparison.

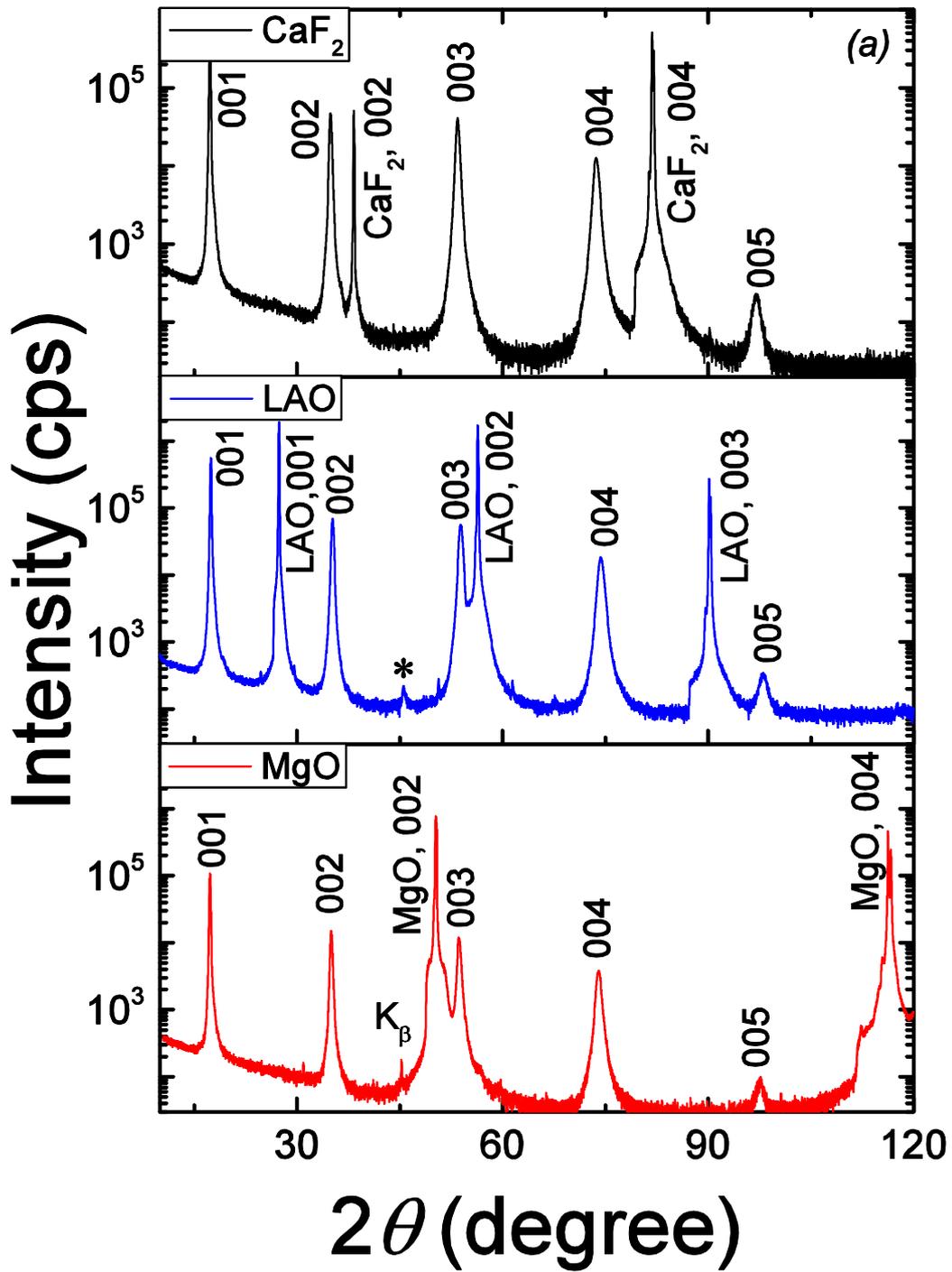

Figure. 1 F. F. Yuan *et al*

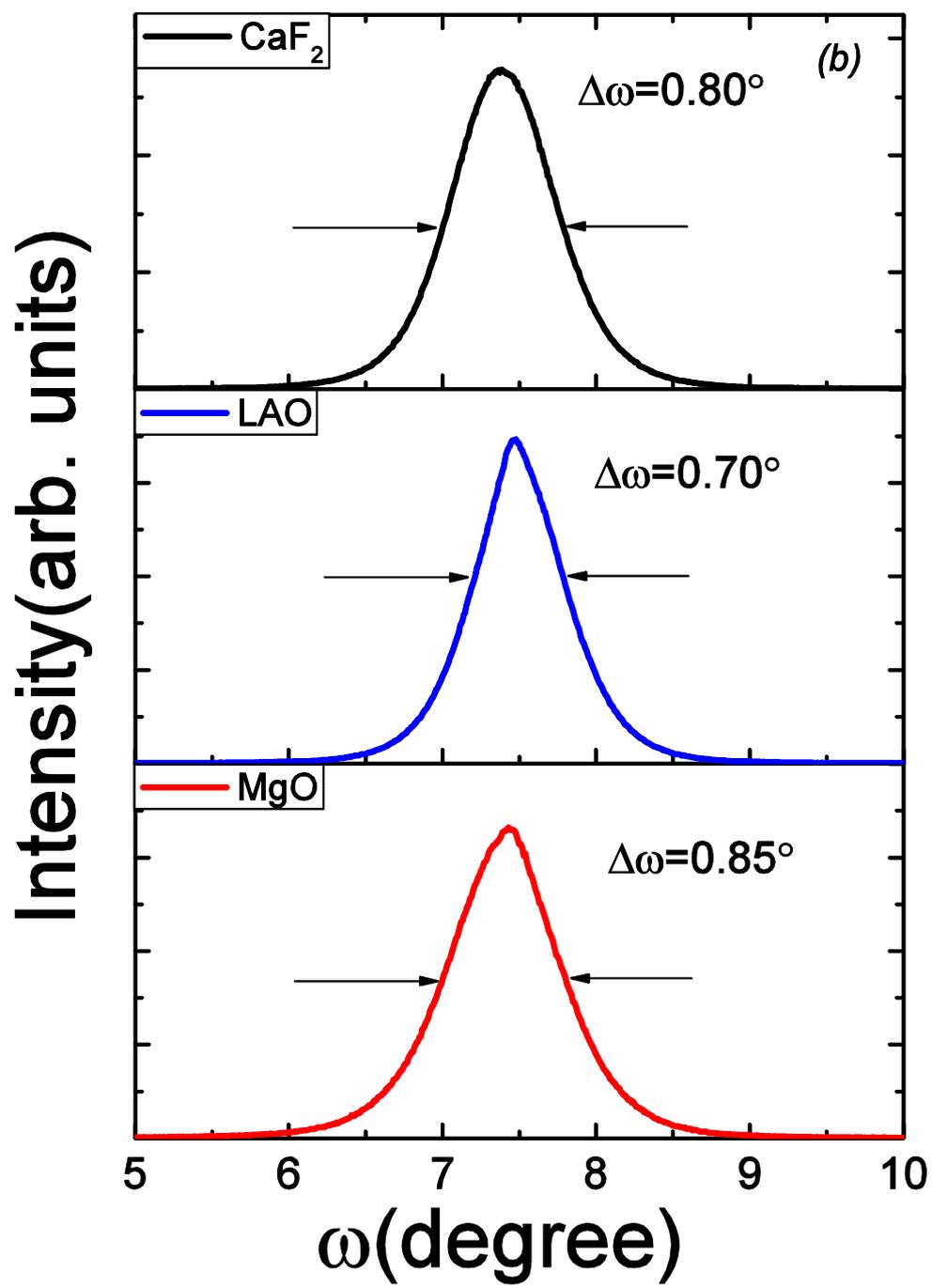

Figure. 1 F. F. Yuan *et al*

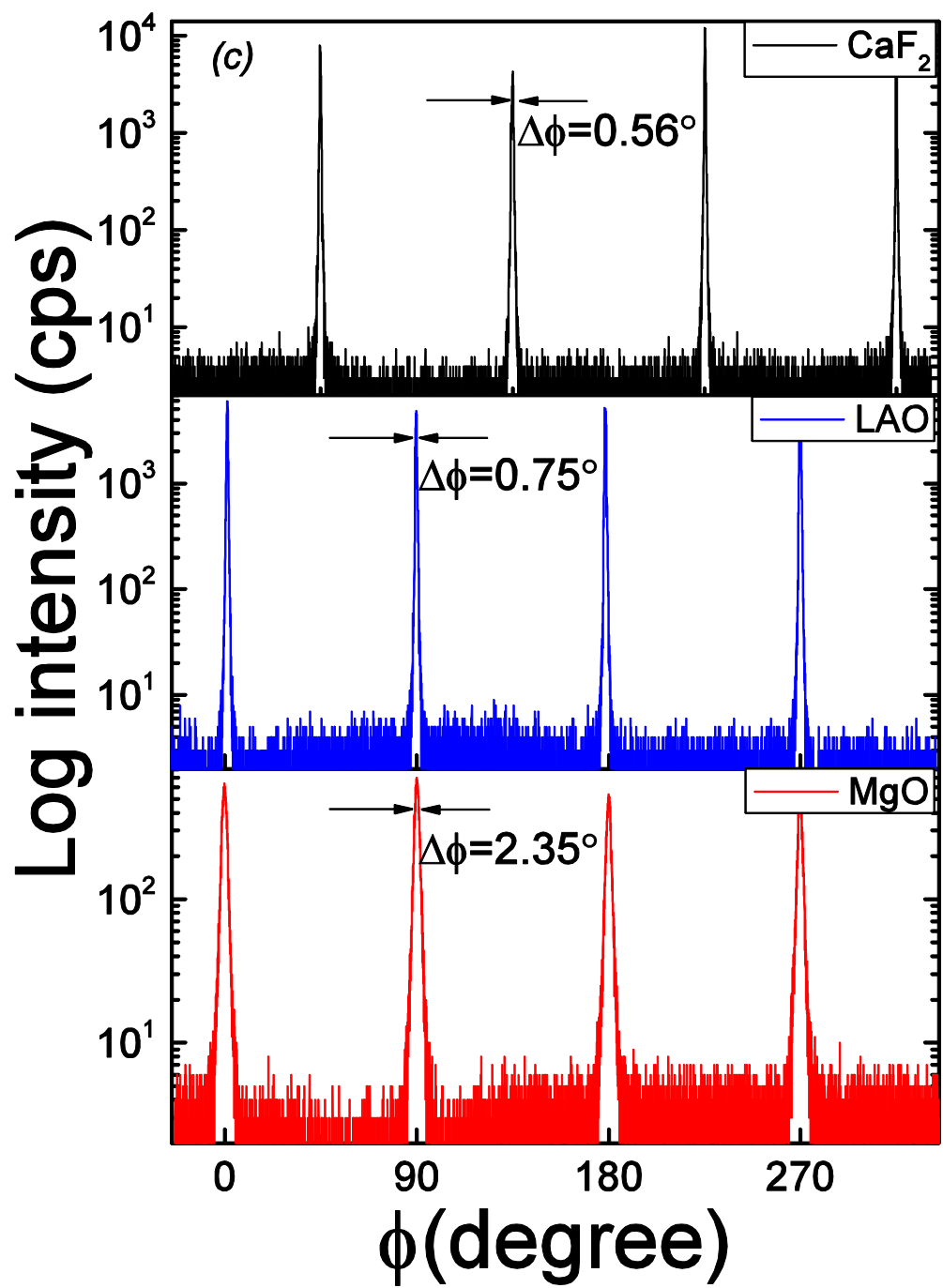

**Figure. 1** F. F. Yuan *et al*

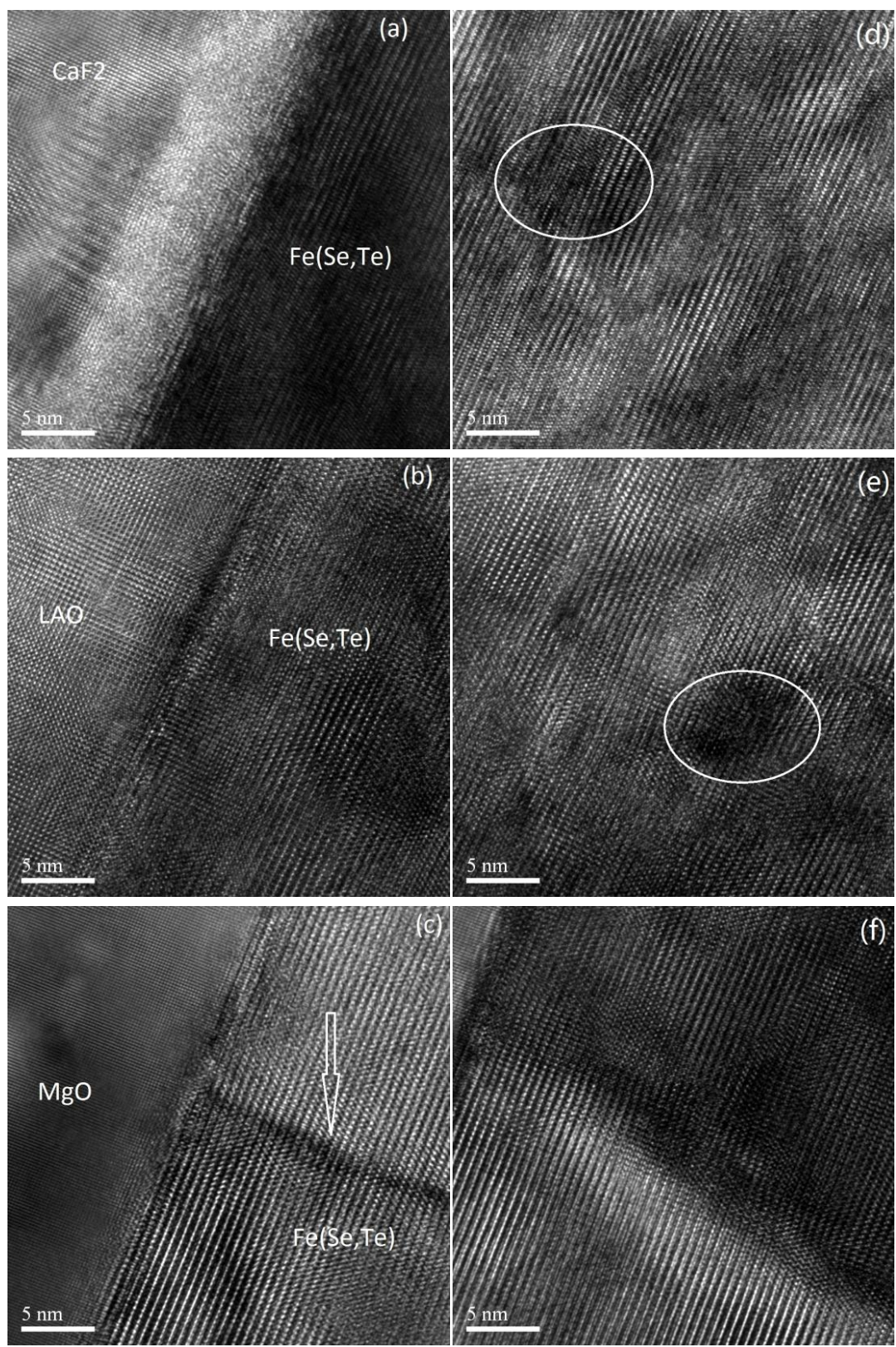

**Figure. 2 F. F. Yuan *et al***

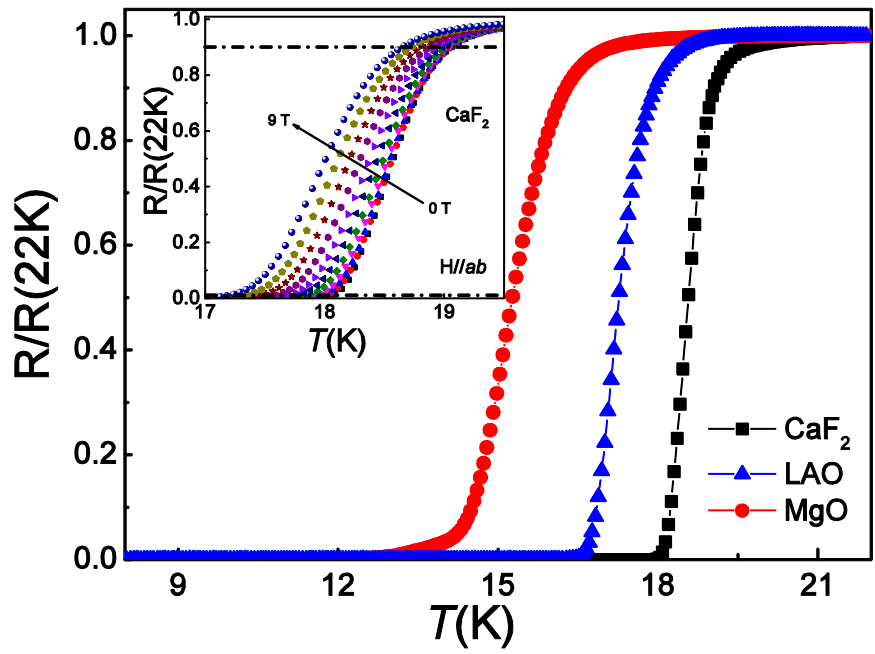

**Figure. 3** F. F. Yuan *et al*

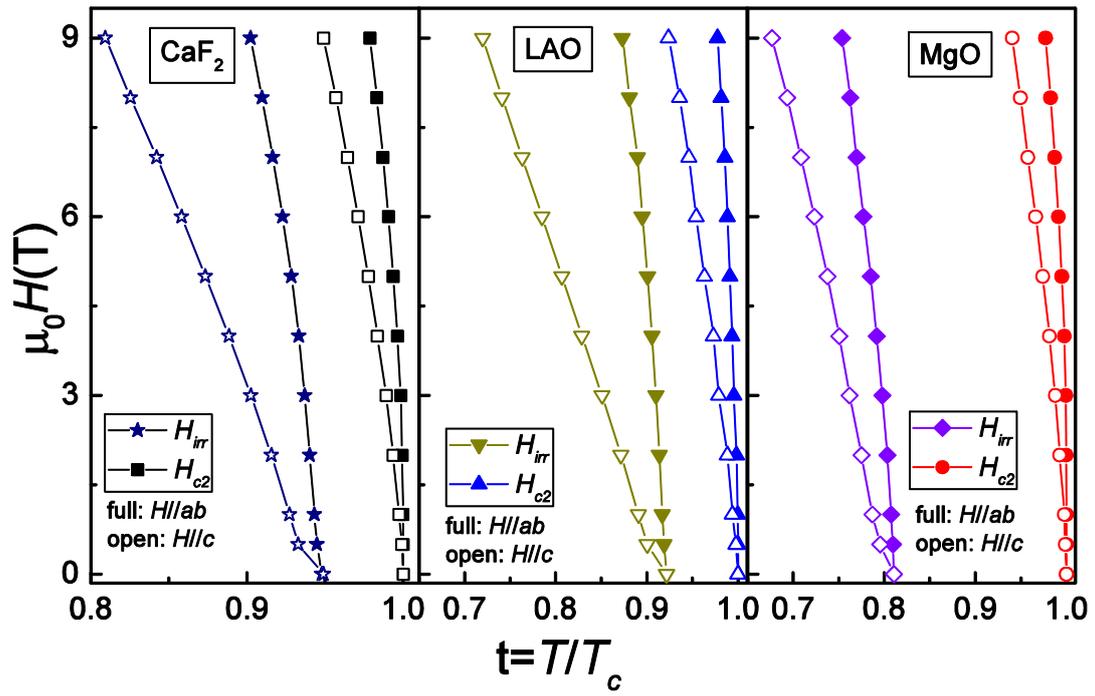

Figure. 4 F. F. Yuan *et al*

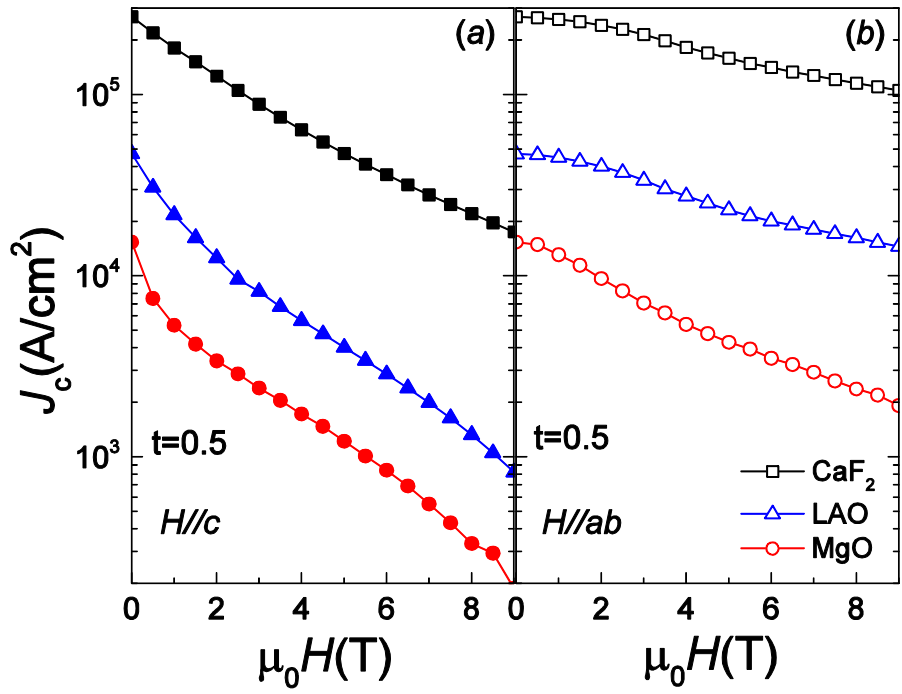

Figure. 5 F. F. Yuan *et al*

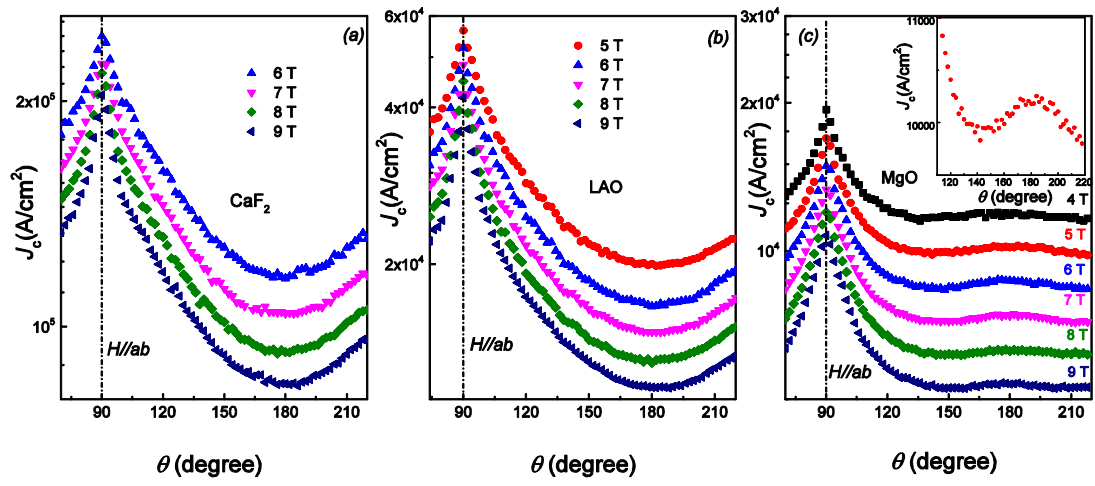

**Figure. 6** F. F. Yuan *et al*

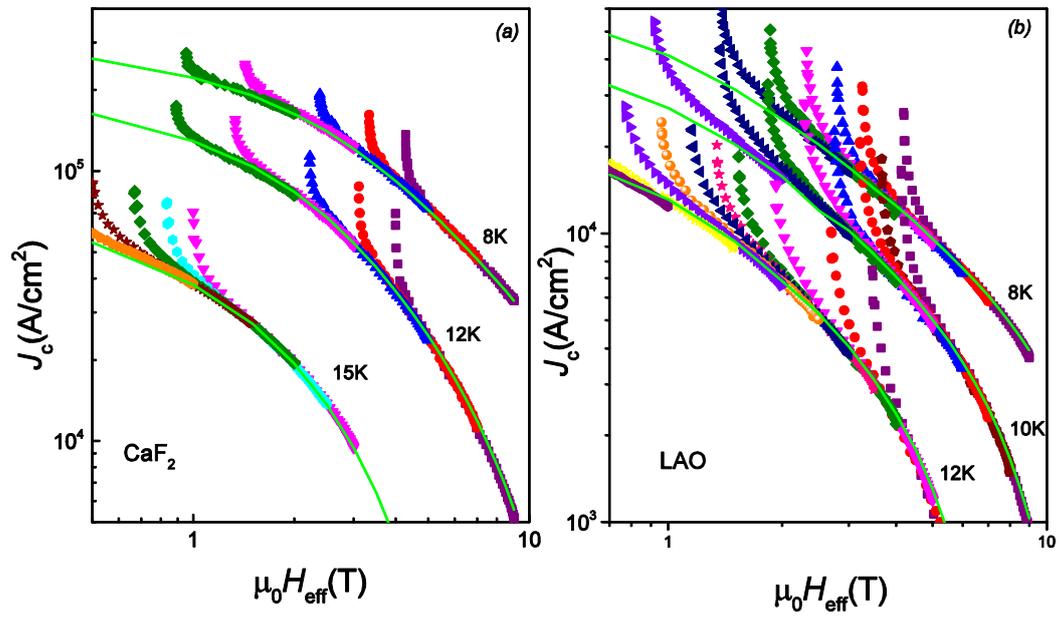

**Figure. 7** F. F. Yuan *et al*

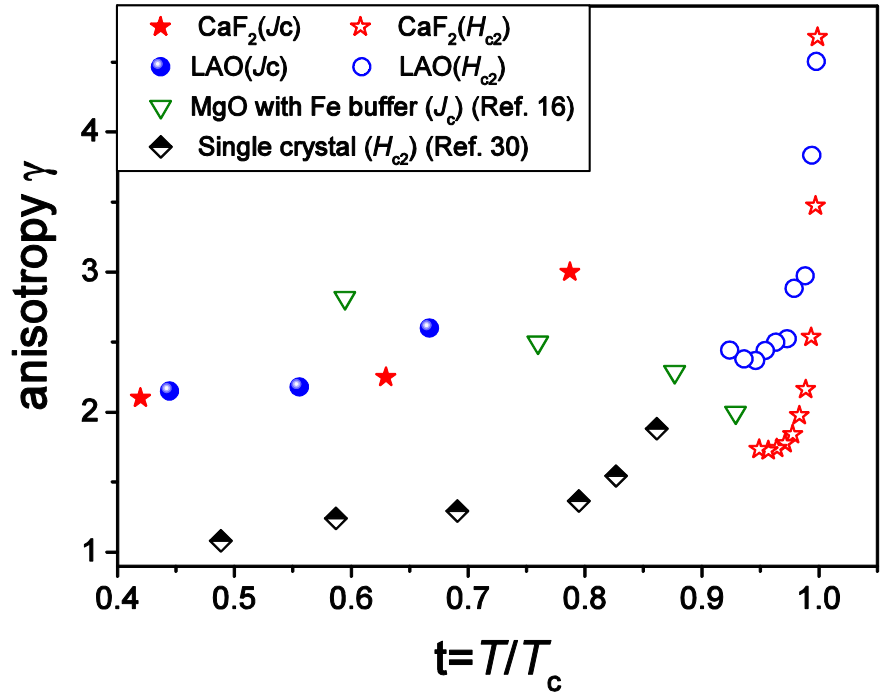

Figure. 7 F. F. Yuan *et al*